\documentclass{PoS}
\def\kms      {\ifmmode{\rm km\,s}^{-1} \else km\,s$^{-1}$\fi}
\def\mujybm{\ifmmode{\rm \mu Jy}\,{\rm beam}^{-1}\else${\rm \mu}$Jy\,beam$^{-1}$\fi}
\def\ltsim{\ifmmode\stackrel{<}{_{\sim}}\else$\stackrel{<}{_{\sim}}$\fi}
\def\gtsim{\ifmmode\stackrel{>}{_{\sim}}\else$\stackrel{>}{_{\sim}}$\fi}
\def\farcs{\hbox{$.\!\!^{\prime\prime}$}}

\def\hour{\hbox{$^{\rm h}$}}

\title{Radio Supernovae}

\ShortTitle{Radio Supernovae}

\author{\speaker{R. J. Beswick}\\
        Jodrell Bank Observatory, The University of Manchester, Lower Withington, Nr. Macclesfield, Cheshire, SK11~9DL\\
        E-mail: \email{Robert.Beswick@manchester.ac.uk}}

\abstract{In this article I will briefly review the current
status of radio observations of nearby supernovae and
their remnants. This review will initially address observations of the radio light curves of nearby
core-collapse supernovae, followed by a more detailed summary of
recent Very Long Baseline Interferometric observations of the
expansion of nearby supernovae and their remnants. These later
sections will concentrate on a few  sources, namely those found in M82, SN1993J, and the recent
supernovae SN2004et. In addition I will discuss the many radio
detections of supernovae found in the highly obscured centres of starburst galaxies,
such as M82 and Arp220, where no optical detections are possible.}

\FullConference{The 8th European VLBI Network Symposium on New Developments in VLBI Science and Technology
                and EVN Users Meeting  \\
		September 26-29 2006\\
		Torun, Poland}

\begin{document}

\section{Introduction}

Radio supernovae and their remnants not only provide a tracer
of the ongoing star-formation within galaxies but also via radio light
curve monitoring and VLBI imaging allow the physics of the surrounding
circumstellar and interstellar media to be investigated. As such the
study of supernovae at radio wavelengths is complimentary to
observations made at other wavelengths, whilst in all but the
closest supernovae events, such as SN1987A, radio observations
and in particular VLBI observations, provide the only method by which
the structural evolution of supernovae can be directly imaged.

In many nearby galaxies with high star-formation rates, such as M82
and Arp220 numerous compact radio sources are detected
\cite{unger84,kronberg75,muxlow94,kronberg85,smith98}. In these highly
dust obscured regions, which incidentally is were a large amount of the
Universe's star-formation is occurring, the extinction free free nature
of radio observations mean that they are the only method by which
these regions can be studied. With the high angular resolution and
high sensitivity of radio interferometric observations it is possible
to detect, resolve and characterise the nature of these compact
sources. For example, in M82 approaching 50 compact objects are
detected at centremetric wavelengths, the majority of these sources are
identified as supernova remnants, the remaining $\sim \frac{1}{3}$ are
compact H{\sc ii} regions \cite{mcdonald02}. Such sources as M82 and Arp220 provide an
illustrative example of why radio observations are a vital addition
to the numerous studies of supernovae at other wavelengths. In these
two cases, as well as several others (e.g. IIIZw35
\cite{pihlstrom01}, Arp299 \cite{neff04}, Mrk 273
\cite{carilli00,bondi05}, NGC6240 \cite{gallimore04} etc), radio
observations are the only way in which individual supernovae and
their consequent remnants can be directly identified.  Without such radio observations many supernovae within the highly obscured
centres of galaxies may have been overlooked.

\section{Optical classifications of supernovae and optically detected number counts}

\subsection{Supernovae taxonomy}

\begin{figure}
	\centering
		\includegraphics[width=0.80\textwidth]{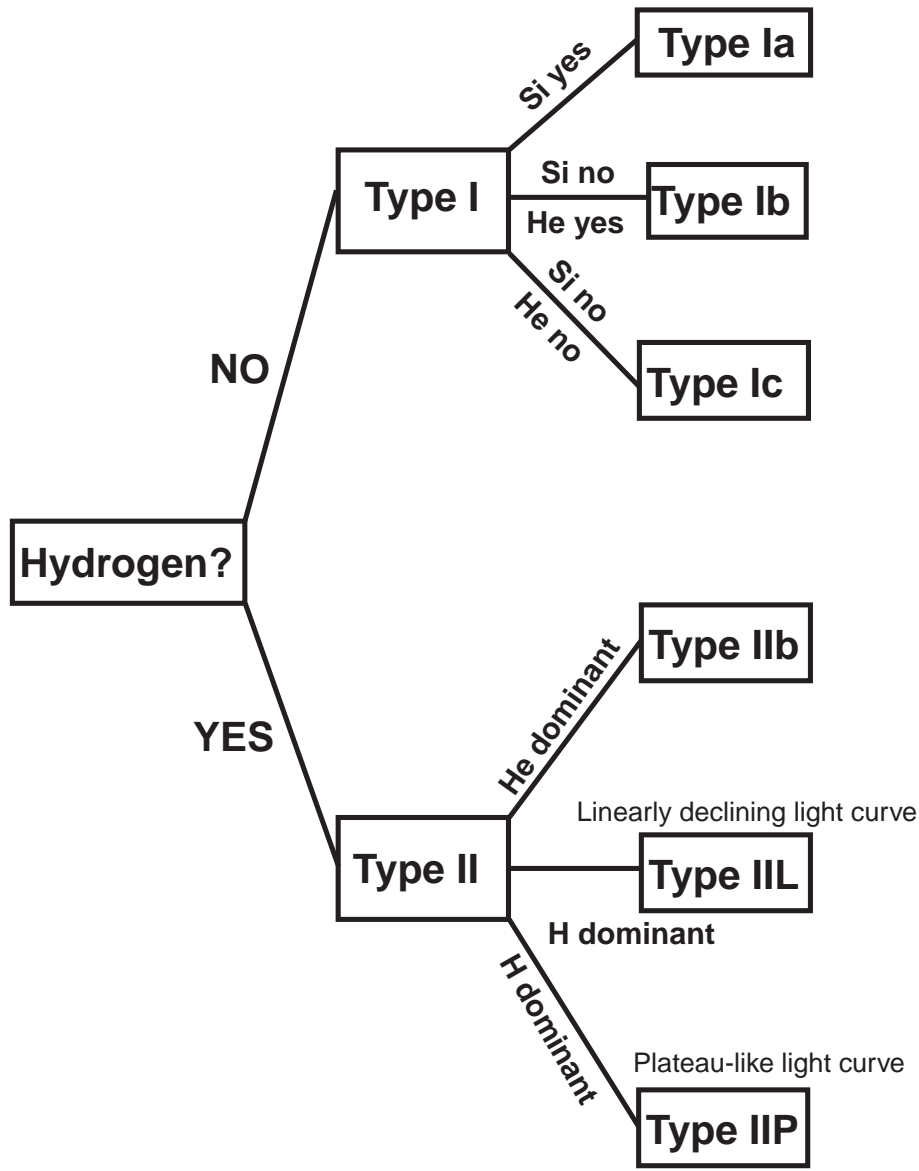}
	\caption{Schematic representation of the major optical classifications of supernovae.}
	\label{fig1}
\end{figure}

Whilst this short review will exclusively concentrate upon radio
observations and in particular high resolution observations it is useful to briefly recap the optical classifications
and definitions of supernovae. Supernovae are optically classified
into Type-II or Type-I sources depending on the detection or
non-detection of hydrogen emission lines in early spectra, respectively. These
fundamental categories are often further sub-divided. In the case of
Type-I supernovae these are traditionally split into Ia, Ib
and Ic on the basis of their optical emission lines. Whilst Type-II
supernovae are similarly classified into various groups on the basis
of their optical spectra and light curve shape. These classes
and their subsequent sub-divisions are schematically shown for the most
common cases in Fig.\,\ref{fig1}. These classifications can
be related to the physical mechanisms undergone within the supernova
and in the star's pre-supernova phase. 

In general, Type-Ia supernovae
are thought to originate from thermonuclear deflagration of material
accreted onto the surface of a white dwarf companion of a more massive
giant star (see \cite{leibundgut00,branch98} and references
therein). As such these Type-Ia supernova are not core collapse in
origin and are analogous to galactic nova sources, such as RS Ophiuchi
which is thought may eventually become a Type-Ia supernova (see
\cite{obrien06a,obrien06b,sokoloski06}). Type-Ia supernovae are often
considered as
relatively homgeneous events in the optical with similar luminosities
and spectral evolution, hence have commonly been used as cosmic
distance indicators. However at the detection limit of current radio
interferometers no radio emission has yet been detected from any
extragalactic Type-Ia supernovae despite numerous deep searches
\cite{panagia06}.

Both Type-1b/c and Type-II supernovae originate from the core collapse
of massive stars rather than the thermonuclear explosions of a low
mass star that is the origin of Type-Ia SNe. These sources are
sub-classifed due to the presence or absence of optical lines
in the early spectra and their optical light curves. In some cases the spectra of a supernova has been seen to evolve
between these classes, a noteable example is the case of SN1993J
which evolved from Type-II to
Type-Ib/c within its first few few weeks. In general Type-Ib SNe show strong
He\,{\sc i} absorption, whereas Type-Ic show only weak He\,{\sc i}
absorption lines. Between these two classes there are no clear
differences observed at radio wavelengths so I will consequently refer to
these sources as Type-Ib/c sources. Other core collapse SNe which show hydrogen
lines in their early spectra are Type-II sources. These are commonly
divided into Type-IIL and Type-IIP supernovae depending of the shape
of their optical light curves. Type-IIL SNe show an approximately
linearly declining optical light curve whereas Type-IIP sources
display a plateau-like optical light curve. The radio SN1979C and
SN1987A provide examples of each of these sub-classes respectively. At
radio wavelengths several examples of each of theses core collapse
supernovae events have been detected (see \cite{weiler02} and
references therein).

\subsection{Detections of optical and radio supernovae}

At present approximately 3 $\rightarrow$ 4 hundred extragalactic
supernovae are reported each year. Almost all of these sources are
detected at optical wavelengths with the majority of these being
reported by professional large scale supernovae search programmes,
such as KAIT \cite{filippenko01}, however some are still discovered by
amateur astronomers.  At radio wavelengths relatively few of
these optically identified sources are ever detected. To date well in excess of 100
supernovae have lower limits to their radio emission established
whilst only several dozen
sources out of the thousands of optical supernovae have been detected and had detailed observations of their radio
evolution made. Of these radio detected supernovae \emph{all} are core
collapse sources (Type-Ib/c or Type-II). Of the radio detections so
far made the following broad characteristics have been established:

\begin{itemize}

\item{The radio emission from Type-Ia sources falls below the current
sensitivity limit of radio interferometers such as the VLA.}

\item{Type-Ib/c supernovae can be radio luminous. They have rapid
turn-ons/turn-offs at radio frequencies, peaking at centemetric
wavelengths on comparable timescales to their optical emission peak. Their
radio spectra is steep ($\alpha\ltsim1$ [S$\propto\nu^{-\alpha}$]).}
\item{The radio emission from Type-II supernovae shows very varied
properties, including a wide variations in luminosity and timescales. The spectral
indices of Type-II SNe have a tendency to be flatter  and they tend to
have a slower turn-on and can be more long lived than Type-Ib/c sources.}
\end{itemize}

For further details on the radio properites of SNe please see some of
the many excellent and comprehensive reviews, for example Weiler {\it et al.} \cite{weiler02}.

\begin{figure}[htbp]
	\centering
		\includegraphics[width=.8\textwidth]{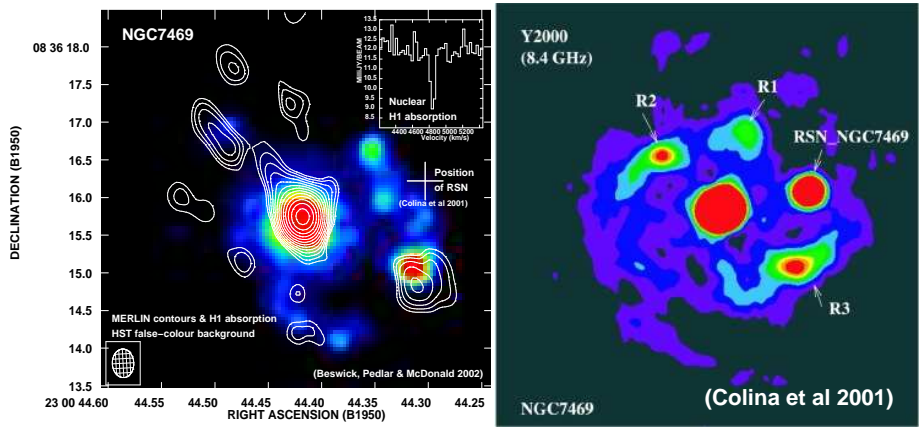}
	\caption{Before and after radio images of the centre of the
nearby Seyfert galaxy NGC7469. In the left-hand image a MERLIN radio
contoured image is shown overlaid upon a {\it HST} FTC2 false colour
image (from Beswick, Pedlar \& McDonald 2002 \cite{beswick02}). This
radio image was taken before SN2000ft was detected. The right-hand
image show an 8.4GHz VLA image of the same region. The new bright
source to the west of the nucleus, situated in the well-known
cirumnuclear starburst ring is interpreted as a bright radio
supernova \cite{colina01,alberdi06}.}
	\label{ngc7469}
\end{figure}

Whilst these charateristics are established for the optically
identified and classified supernovae that have been studied in detail
at radio wavelengths there are many compact radio sources, many of
which are thought to be supernovae, of these sources detected in the centres
of dust obscured galaxies a large fraction are never optically
classified. It is assumed that many of these these are massive core
collapse supernovae, extreme examples of these sources may be the
those found in the centre of Arp220
\cite{smith98,rovilos05,lonsdale06,parra06a,parra06b} or the luminous radio supernovae detected in
NGC7469 \cite{colina01,alberdi06} (see Fig\,\ref{ngc7469}).

\section{Detecting and monitoring supernovae at radio wavelengths}

The multiwavelength detection and monitoring of the radio emission from supernova events provides a wealth of information about the
density and structure of the circumstellar medium (CSM). From the
modelling of these radio 'light' curves it is possible to
derive evidence about the clumpiness or filamentary nature of the
stellar wind prior to the star becoming a supernova, the mass-loss history of the
progenitor star, possible binary companions of the progentor as well as
constraining the explosion date for the supernova event
(see Weiler {\it et al.} \cite{weiler02}  for a review of this work and references therein). 

The radio 'light' curve of a supernova goes through an initial rise,
following the explosion, as the material surrounding the star becomes
less dense and opaque as it expands. This material becomes more
transparent at higher frequencies first, so the explosion becomes
visible at higher frequencies before the light curve rises at lower
frequencies. As this expanding envelope grows larger it
cools. At a certain point the material becomes cool enough that it no
longer radiates and at optical wavelengths the source fades. For core
collapse radio supernovae the radio emission is thought to originate
from an approximately spherical expanding shell just behind the
blast-wave interaction with the CSM. This radio emission is
non-thermal synchrotron emission originating from the acceleration of free
electrons in the enhanced magnetic field arising from the shock wave
interaction of the blast-wave and the CSM. As a consequence the
surrounding ambient conditions of the CSM dictate the evolution of the
supernovae radio emission. 

\begin{figure}[htbp]
	\centering
		\includegraphics[width=1.0\textwidth]{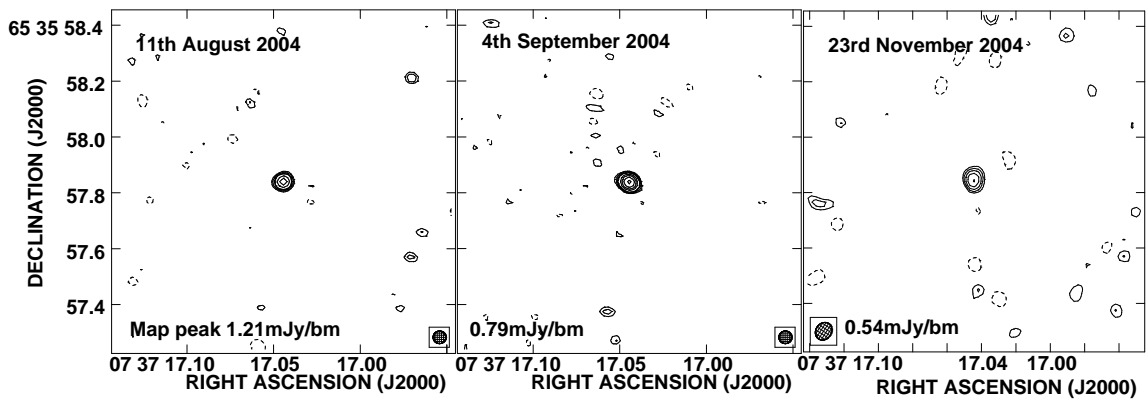}
	\caption{Images of the faint radio emission from the of the optically bright nearby
supernova SN 2004dj in NGC2403 from Beswick {\it et
al.}\cite{beswick05}.}
	\label{2004dj}
\end{figure}

\subsection{One nearby example: The Type-IIP Supernova 2004dj in NGC2403}

The detection of SN 2004dj, visually on 31 July 2004 \cite{nakano04} and
its detection at radio wavelengths using the VLA at 8.4 GHz on 2 August \cite{stockdale04},
provided an ideal opportunity to study a nearby supernova and its
evolution at radio wavelengths. Peaking at magnitude 11.2, SN\,2004dj
was the brightest optical supernova for several years.  Shortly after
the initial detection it was reported that the spectrum of SN\,2004dj
showed features typical of Type II-P supernovae \cite{patat04}. Type
II-P supernovae are believed to originate via core-collapse in
hydrogen rich, massive stars.  

\begin{figure}[htbp]
	\centering
		\includegraphics[width=0.70\textwidth]{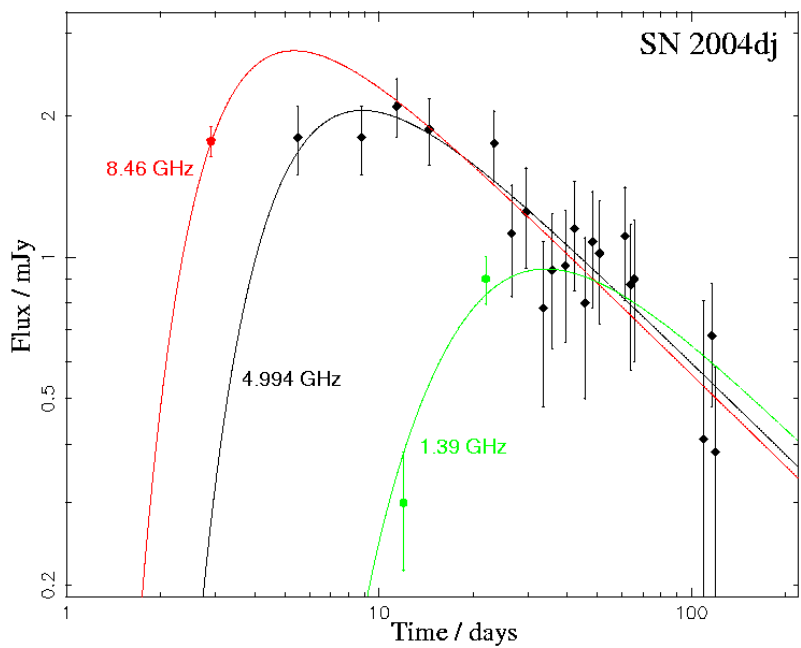}
	\caption{The early radio light curve of the bright nearby
supernova SN 2004dj in NGC2403 from \cite{beswick05,argo05,argo06}.}
	\label{2004djfit}
\end{figure}

In the radio SN2004dj was detected only a few days following the
optical discovery. MERLIN observations using a subset of the array began in early August 2004 and 
continued into early October, followed by imaging runs using the full array in 
November and December 2004 (Fig.\,\ref{2004dj}).  This allowed a detailed 5\,GHz light curve to be 
determined (Fig. \ref{2004djfit}).

These MERLIN observations allowed the position of the source to be determined 
to an accuracy of better than 50\,mas \cite{argo04b}, coincident with the 
optical \cite{nakano04} and $Chandra$ X-ray \cite{pooley04} positions, and 
the star cluster n2403-2866 \cite{larsen99}.  This illustrates the usefulness 
of high resolution observations, for example using MERLIN, for this kind of program as, at the time, the VLA was in the most 
compact D configuration.  Although 2004dj was observed with the VLA 
\cite{stockdale04}, the measured position was affected by extended emission 
from the galaxy and lies $\sim$1\farcs2 from other measurements.

This SNe was optically classified as Type II-P \cite{patat04}, a relatively 
common type of supernova optically, but rarely detected at radio wavelengths.  
In fact, prior to 2004dj, the only two Type II-P SNe detected by radio 
telescopes were SN 1999em \cite{pooley02} for which no light curve was 
established, and the well-observed SN 1987A \cite{turtle87}, both of which 
were also relatively weak radio emitters. With the exception of the
LMC supernova 1987A, observations of SN2004dj provide the only
detailed information regarding the radio 'light' curve of this class
of core collapse supernova.

\section{Radio imaging of nearby Supernovae and Supernova remnants}
\subsection{Supernova remnants in M82}

As has previously been mentioned the central starburst region within
the nearby irregular galaxy M82 contains a wealth of compact sources
that are the consequence of the massive and ongoing star-formation
within the centre of this source. Early high resolution radio
observations by Unger {\it et al.} \cite{unger84} and Kronberg {\it et
al.} \cite{kronberg75,kronberg85} identified these sources as
population of supernovae and supernova remnants. Latter more detailed
work (\cite{muxlow94,mcdonald02} ) have shown that these
sources are a mixture of supernova remnants and compact H{\sc ii}
regions.  

\begin{figure}
	\centering
		\includegraphics[width=.80\textwidth,angle=0]  {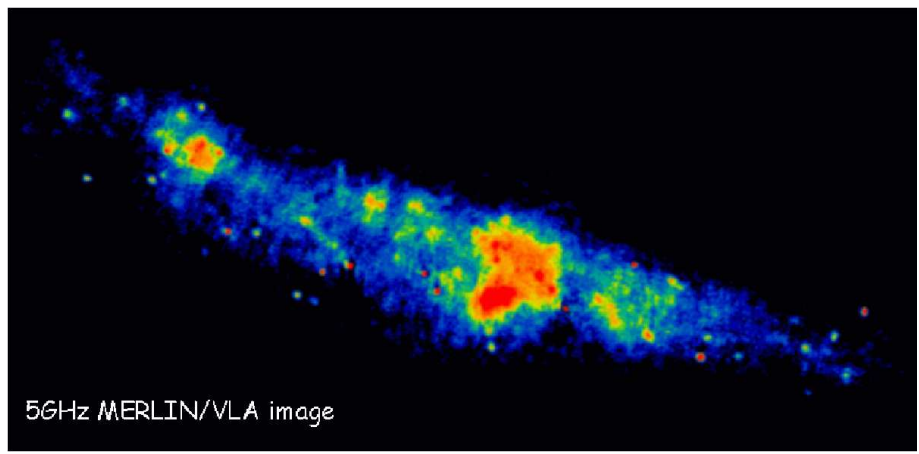}
	\caption{False colour combined MERLIN+VLA 5GHz image of M82
showing numerous compact radio sources embedded within the diffuse radio
continuum. MERLIN and VLBI observations resolve all of these compact
sources (Courtesy of Tom Muxlow).}
	\label{m825ghz}
\end{figure}

Over the $\sim$20 years high resolution VLBI and MERLIN observations
have been used to resolve and monitor the structural evolution of
these sources in great detail
\cite{wilkinson84,bartel87,muxlow94,trotman96,pedlar99,mcdonald01,muxlow05,beswick06,fenech06} (see Fig.\, \ref{m825ghz}). 

\begin{figure}
	\centering
		\includegraphics[width=.30\textwidth,angle=270]  {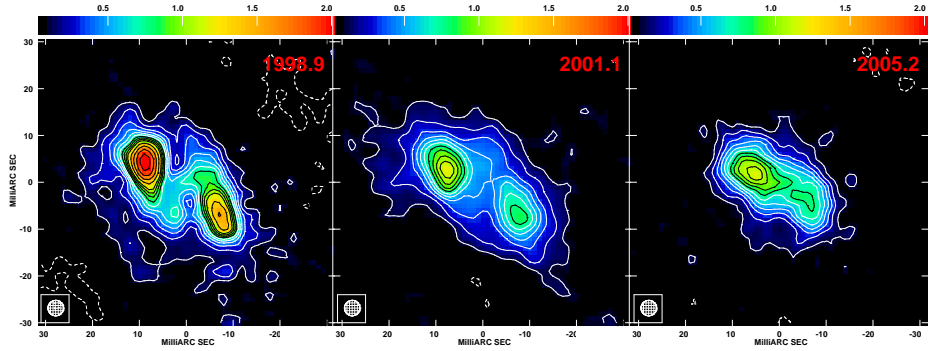}
	\caption{Contour images of the compact radio source 41.95+575
from the global VLBI epochs observed on 1998 November 28, 2001
February 23 and 2005 March 3rd. Each images have been convolved with a
3.3-mas circular beam. The three epochs are contoured with linear
multiples ( 1, 1, 2, . . . 10)$\times$0.21,  0.11 and
0.11\,mJy\,beam$^{-1}$ respectively. Images are taken from \cite{mcdonald01,beswick06,fenech06}.}
	\label{41.9}
\end{figure}

The two most compact sources in M82 (41.95+475 and 43.31+592) have been
resolved in detail using a series of global VLBI observations over the
last 20 years \cite{pedlar99,mcdonald01,beswick06,fenech06}. Three
18cm global VLBI images of the most compact source in M82 are shown in
Fig.\,\ref{41.9}. This series of 4\,mas angular resolution images
observed in 1998, 2001 and 2005  show evolving radio
structure of 41.94+475.  Although initially considered a compact
radio supernova remnant, similar to many of the other sources in M82,
41.9+475 displays a distinctly different radio structure. The compact
VLBI structure of this source appears to be bi-polar in nature with
two 'hot-spots'. These two components appear to be moving apart quite
slowly with expansion velocities in the range of only $\sim$1000 to
2000\,km\,s$^{-1}$ between 1998 and 2001 although the most recent
epoch shows considerable further structural evolution making a
definitive velocity calculation hard to achieve (see
\cite{fenech06}). Whilst it is clear that 41.95+475 is evolving both
in its radio flux density and milliarcsecond radio structure it is
still unclear as to the true nature of this source. It has recently
been speculated that this source is not a true radio supernova event
but something more exotic, such as the remnant of an off-axis
GRB (see \cite{muxlow05,beswick06,fenech06}).

\begin{figure}
	\centering
		\includegraphics[width=1.00\textwidth]{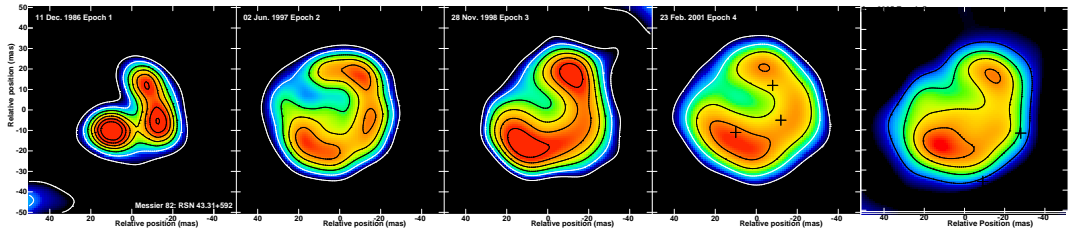}
	\caption{Contour images from all five VLBI epochs of the
compact RSN 43.31+592. From left to right these observations were made on 1986 Dec 11th,
1997, June 2nd, 1998 Nov 28th, 2001 Feb 23rd and 2005 March 03 respectively, All five epochs have been
convolved with a circular beamsize of 15 mas, to match the earlier
EVN-only epochs in 1986 and 1997. All five images are contoured with
1, 1, 2, 3, 4, 5, 6, 7, 8, 9 and 10$\times$ 0.35\,mJy\,beam$^{-1}$. On
the fourth, three crosses are plotted to show the relative
Gaussian-fitted positions of the three components in epoch 1. Images are taken from \cite{mcdonald01,beswick06,fenech06}.}
	\label{43.3}
\end{figure} 

Unlike 41.95+475 the second most compact radio source in M82,
43.31+592 resembles the classical shape for a radio supernova
remnant. As can be seen in Fig.\,\ref{43.3} 43.31+47 shows a regular
ring-like structure with breakout region toward the north-eastern
corner. Over the 20\,years of monitoring the radio structure and size of this
remnant have been studied in detail. These observations have shown
that this source is in near free expansion with a velocity of between
9000 and 11000\,km\,s$^{-1}$
\cite{pedlar99,mcdonald01,beswick06}. This value is close agreement
with other radio expansion measurements of this source using MERLIN
from Muxlow {\it et al.} \cite{muxlow05b}. Both of these sources along
with several of the other compact radio supernova remnants in M82 are
discussed in more detail elsewhere in this proceedings \cite{fenech06}.

\subsection{SN1993J in M81}

\begin{figure}
	\centering
		\includegraphics[width=0.35\textwidth]{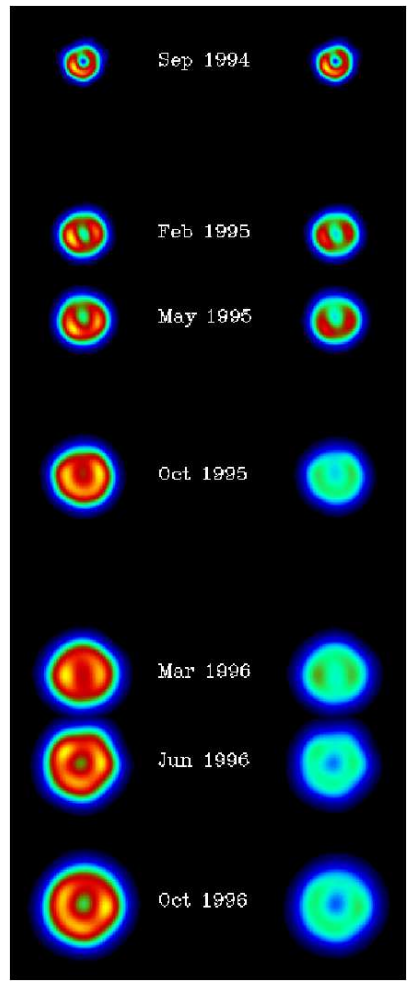}
	\caption{Global VLBI images showing the first few years of the
expansion of SN 1993J in M81 \cite{marcaide94a,marcaide95}.}
	\label{1993J-1}
\end{figure}

Of all supernovae observed with high resolution VLBI, SN1993J in M81
has been subject to the most detailed scrutiny. SN1993J exploded on
March 28th 1993 and was first detected at radio wavelengths just a few
days later at the beginning of April 1993
\cite{weiler93,pooley93}. By virtue of its closeness, (M81 is just 3.6\,Mpc
away) its brightness (at the time the nearest northern hemisphere
supernova since 1937) and its high northern declination
made this source an ideal target for VLBI observations. The source was
first detected using VLBI on 1993 April 25th, these observations
established the position of the supernova to an accuracy of a few
milliarcsec \cite{marcaide93a,marcaide93b}.

Early size estimates of SN1993J were established approximately 30 days
following the supernova explosion implying an angular size of 0.28$\pm$0.06\,mas on 1993 May
1st \cite{marcaide94}. Following this a series of VLBI measurements
of the sources structure and expansion have been made
\cite{bartel94,marcaide94a,marcaide95}. The initial expansion rate of
the radio emission was $\sim2.4\mu$arcsec\,day$^{-1}$. Subsequent VLBI imaging over more than a
decade has resolved SN1993J into a self-similarly expanding radio
shell (see Fig.\,\ref{1993J-1}) with a recorded deceleration parameter
of between 0.79 and 0.9 ($\theta\propto$t$^m$). After approximately day
1550 an increased deceleration was observed at 6cm. 

\begin{figure}
	\centering
		\includegraphics[width=0.50\textwidth]{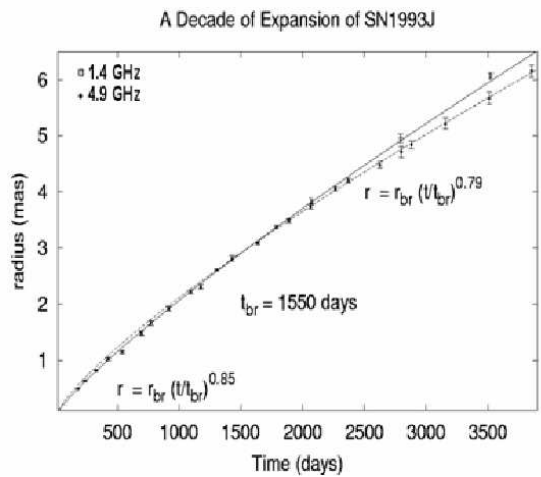}
	\caption{A decde of expansion of SN1993J. Plot of size
measurements derived from VLBI observations for the expanding shell of
SN1993J showing the expansion to be decelerating \cite{marcaide06}.}
	\label{1993J-decel}
\end{figure}

Very recent results from Marcaide {\it et al.} \cite{marcaide06}
conclude that the expansion of SN1993J is actually in self-similar, even
if at 6cm there appears to be increased decelerate after day $\sim$1500 (see
Fig\,\ref{1993J-decel}). This kink in the expansion
curve can be explained by a combination of the partial lifting of the
absorption of the emission from the back side of the shell behind the ionized
ejecta and a pronounced radial drop of the magnetic field biasing the
measurement of the ring size. 

Due to the immense detail in which it has been possible to study
SN1993J at radio wavelengths numerous articles have been written (many
of which are not mentioned here). For a more detailed precis of the
impact of the observations of SN1993J please refer to the numerous
articles published in \emph{Cosmic Explosions: On the tenth
anniversary of SN1993J} IAU colloquium192, Eds. Marcaide \& Weiler.

\subsection{SN2004et in NGC6946}

\begin{figure}
	\centering
		\includegraphics[width=0.50\textwidth]{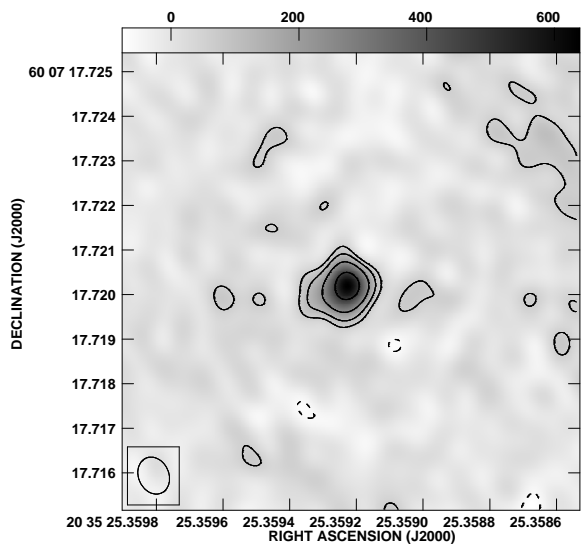}
	\caption{Global VLBI image of SN2004et on day 153 (from
Marti-Vidal {\it et al.} \cite{ivan06}). Map peak flux is
0.639\,mJy/bm and convolved is with a 0.86$\times$0.68\,mas beam.}
	\label{2004et}
\end{figure}

In September 2004 a bright supernova was discovered, SN 2004et 
\cite{zwitter04}.  The host galaxy, NGC\,6946, is a nearby active
starburst galaxy which has hosted many historical supernovae.
Initially it was hoped that this nearby, optically luminous, Type-II
supernova would become luminous at radio wavelengths providing an
opportunity similar to that of SN1993J to observe its radio structural
evolution in great detail. However following early radio monitoring using both MERLIN and the VLA
\cite{beswick04,argo06} it became apparent that the radio emission of
SN2004et would not reach a high luminosity. The early light curve
monitoring of this source recorded a peak flux density of
$\sim$2.5\,mJy\,beam$^{-1}$ at 5GHz at an age of approximately 40
days.  However, a first epoch of sensitive 8.4\,GHz global VLBI
observations were made by Marti-Vidal {\it et al} \cite{ivan06} on
2005 February 20th (day 153). These observations detected and
marginally resolved SN2004et (Fig.\,\ref{2004et}). 

The radio structure of SN2004et shows a bipolar geometry which has
been interpreted as either coming from a shell with two hot spots or
an ejecta that is expanding anisotropically. Favouring the former of
these interpretations an expansion velocity of
16000$\pm$2000\,km\,s$^{-1}$ is derived. This velocity is comparable
to the early expansion velocities recorded for other similar radio supernovae.

\subsection{Optically undetected Radio supernovae}
 
\begin{figure}
	\centering
		\includegraphics[width=.55\textwidth,angle=270]{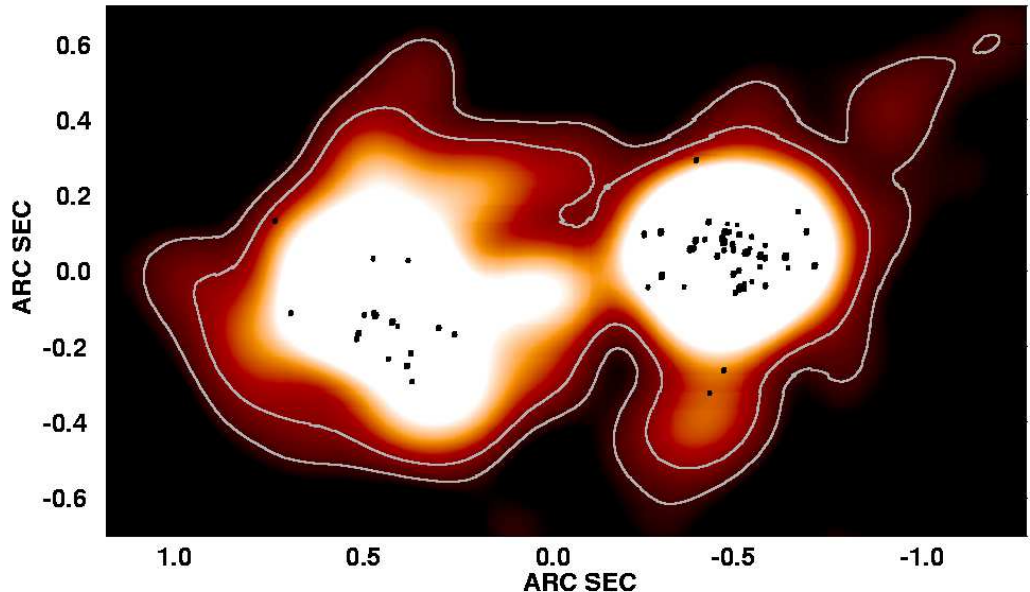}
	\caption{Extremely deep 1.6\,GHz global VLBI image of the
centre of Arp220. The small black dots show the high resolution VLBI
sources (Lonsdale {\it et al.} \cite{lonsdale06}). This VLBI image is overlaid upon a 'lower' resolution MERLIN 1.4GHz image of the central region of Arp220 from Mundell, Ferruit \& Pedlar \cite{mundell01}.}
	\label{arp220}
\end{figure}

In general this review has concentrated on a few examples of
extragalctic radio
supernovae and supernova remnants that have been well studied at radio
wavelengths. However as eluded to earlier there are several
star-forming galaxies
where radio observations, and in particular VLBI observations, are now
begining to reveal large numbers of compact radio components which are
thought to be powerful radio supernovae or supernova remnants. One
example of this is the nearby ultra-luminous infrared galaxy
Arp220. Within the centre of Arp220 $\sim$50 compact
milliarcesecond-scale radio components have been detected by Smith
{\it et al} \cite{smith98}, Rovilos {\it et al.} \cite{rovilos05} and
Lonsdale {\it et al.} \cite{lonsdale06}. As can be seen in
Fig.\,\ref{arp220} these numerous compact components are burried within
the central starburst of Arp220 and have required the extremely high
resolution and high sensitivity of current VLBI arrays in order to
detect them. More recently, and reported in this meeting by Parra {\it
et al.} \cite{parra06a,parra06b}, multi-frequency VLBI observations along with
long-term monitoring \cite{thrall06} of these sources has begun to
confirm them to be a combination of powerful radio supernovae and
supernova remnants. 

High resolution radio observations of objects, such as Arp220 and M82,
remain one of the few ways in which the supernovae and remnants with
these dust, highly obscured starburst galaxies can be
explored; further underlining the necessity of radio observations of supernovae.

\section{Summary and future work}

Whilst this, somewhat incomplete, review of radio supernova and their
remnants has made some efforts to report a few of the highlights of
radio, and in particular VLBI observations, of supernovae. It should be noted that
several other supernovae have also been extensively observed. In particular
SN1973C \cite{bartel85,bartel03}, SN1986J
\cite{bietenholz02,bartel91} and 1987A. Each of these sources has been
subject massively important individual studies, for which there is not
space here to fully do them justice.

In summary, whilst there has been a vast amount learnt about
radio supernovae, radio supernova remnants, the CSM surrounding radio
supernovae and their progenitors via both radio flux density
monitoring and detailed VLBI imaging, to date only a relatively small
fraction of supernovae have been studied at these wavelengths. It is
clear that only a few supernovae detected optically are radio emitters
at levels that can be studied with current instruments, which naturally limits
the number of objects that can presently be studied in
detail. However, considering vast amounts of information that can be
obtained even from relatively few objects I can only describe the
future of this somewhat explosive subject as luminous! This is
specially true considering the advent of new more sensitive radio arrays, such as the EVLA and
e-MERLIN, along with the more rapid response of high sensitive VLBI
arrays through real-time observations (eVLBI).

\section*{Acknowlegdements}

I would like to thank Phil Diamond, Tom Muxlow, Alan Pedlar, Megan Argo, Danielle
Fenech, Jon Marcaide, Ivan Marti-Vidal and Rodrigeo Parra for their many
invaluable contributions to this review.

\end{document}